\documentclass[aps,prl,twocolumn,10pt]{revtex4-1}

\usepackage{graphicx}

\graphicspath{{./Images/}{./plots/3Sbasic/}{./plots/3SfinalHDR/}{./plots/3SfinalHDR/timeseries/}}

\begin{document}

\title{Picometer-stable hexagonal optical bench to verify LISA phase extraction linearity and precision}

\author{Thomas S. Schwarze}
\email{thomas.schwarze@aei.mpg.de}
\author{Germ{\'a}n Fern{\'a}ndez Barranco}
\author{\mbox{Daniel Penkert}}
\author{Marina Kaufer}
\altaffiliation{Currently with SpaceTech GmbH (STI),
  Seelbachstrasse 13, 88090
  Immenstaad am Bodensee, Germany}
\author{Oliver Gerberding}
\author{Gerhard Heinzel}
\affiliation{Max Planck Institute for Gravitational Physics (Albert
  Einstein Institute), Callinstrasse 38, 30167 Hannover, Germany}
\affiliation{\mbox{Leibniz Universit{\"a}t Hannover, Institut f{\"u}r
  Gravitationsphysik}, \\ Callinstrasse 38, 30167 Hannover, Germany}

\date{\today}

\begin{abstract}
The Laser Interferometer Space Antenna (LISA) and its metrology chain have to fulfill stringent
performance requirements to enable the space-based detection of
gravitational waves. This implies the necessity of performance
verification methods. In particular, the extraction of the interferometric phase,
implemented by a phasemeter, needs to be probed for linearity
and phase noise contributions. This Letter reports on a hexagonal quasimonolithic optical
bench implementing a three-signal test for this purpose. Its
characterization as sufficiently stable down to picometer
levels is presented as well as its usage for a benchmark
phasemeter performance measurement under LISA conditions.
These results make it a candidate for the core of a LISA metrology verification facility.  
\end{abstract}



\maketitle

\textit{Introduction}---The first detections of gravitational waves by the Laser Interferometer
Gravitational-Wave Observatory (LIGO) \cite{Abbott2016first} and
Virgo \cite{Abbott20173D} have opened the window for
gravitational wave astronomy in the Hz and kHz range. Avoiding limitations
by seismic and gravity gradient noise, the Laser Interferometer Space
Antenna (LISA) \cite{Bender1998,Danzmann2017} will offer revolutionary
science with sources only detectable in the mHz regime. Important examples are extreme mass ratio inspirals for
strong-gravity tests of general relativity or massive black hole binaries
at red shifts up to 20 to study their early formation \cite{Danzmann2017}.

LISA consists of three spacecraft (SC) forming a triangular constellation with 2.5
million km arm lengths. It will measure the displacement
between free-falling test masses (TMs) by means of heterodyne laser
interferometry. The latter is split into local TM-SC and 
remote SC-SC displacement measurements.  The targeted band of 0.1 mHz to 1 Hz
will be limited above 3 mHz by the optical metrology ($10 \ \mathrm{pm}
/\sqrt{\mathrm{Hz}}$; with $4.7 \ \mathrm{pm}
/\sqrt{\mathrm{Hz}}$ shot noise), and below by TM stray accelerations ($3 \
\mathrm{fm}\, \mathrm{s}^{-2} /\sqrt{\mathrm{Hz}}$). 

The LISA Pathfinder mission impressively showed the
feasibility of the stringent stray acceleration target \cite{Armano2018} as well as local
interferometry with $35 \ \mathrm{fm}/\sqrt{\mathrm{Hz}}$ precision \cite{Armano2016}.
In addition, the recently launched satellite geodesy mission Gravity Recovery and Climate Experiment Follow-On
(GRACE-FO) carries the first SC-SC interferometer with $80 \ \mathrm{nm}
/\sqrt{\mathrm{Hz}}$ targeted precision \cite{Sheard2012}.  
Like a LISA SC-SC interferometer, it operates at weak-light levels ($\sim$
100 pW). However, LISA is more demanding in
terms of precision and in the aspects explained in the following.

Most importantly, each LISA SC-SC interferometer will exhibit coupling
of the full laser frequency noise due to the SC distances acting as huge interferometer arm
mismatches. Mitigation of this otherwise overwhelming
noise coupling will be performed by a technique called Time Delay
Interferometry (TDI) \cite{Tinto1999,Tinto2014}. Primarily, it
time-shifts and combines SC-SC measurements throughout the
constellation in postprocessing to cancel multiple but delayed occurrences of laser
frequency noise. However, as TDI is performed in postprocessing, the
interferometer phase extraction by a phasemeter has to conserve the essential displacement information
hidden in the much stronger laser frequency noise. Therefore, linearity over
a high dynamic range is needed while maintaining phase fidelity. The latter
is defined by the phasemeter (single-channel) noise contribution
requirement of $1 \ \mathrm{\mu cycle} /\sqrt{\mathrm{Hz}}$ 
or $2 \pi\ \mathrm{\mu rad}/\sqrt{\mathrm{Hz}} $ $(\sim 1 \ \mathrm{pm}
/\sqrt{\mathrm{Hz}}$ for a $1064 \ \mathrm{nm}$ laser) down to 3
mHz with a relaxation due to the dominating
TM stray acceleration below. Hence, with expected master laser stabilities
\cite{Sheard2010}, the required dynamic range in orders of magnitude
is 8 at 1 Hz and rises to 10 below 3 mHz.

Also, the phasemeter needs to cope with heterodyne frequencies of 5-25
MHz and change rates up to $20 \  \mathrm{Hz/s}$.  These values are determined
by the expected Doppler shifts due to SC motion as well as by 
offsets intentionally applied for constellation-wide frequency planning.

As a consequence, the verification of the described phasemeter requirements
is crucial to complete the demonstration of the LISA metrology. A
frequently utilized verification scheme is a split test,
which is based on a differential measurement of identical signals. In the past, various
phasemeters have been reported to show phase noise
performances \cite{Gerberding2015,Mitryk2010,Shaddock2006} below $1 \ \mu
\mathrm{cycle}/\sqrt{\mathrm{Hz}}$ in split tests.
Yet, particularly for LISA, a more elaborate verification scheme is
needed, e.g. a three-signal test, first mentioned in
\cite{Shaddock2006}. Applying nonidentical signals to the phasemeter
channels, it reveals noise sources that in a split test cancel out undetected as
common-mode. Most importantly, it also allows the
test of phase extraction linearity, as shown later. However, conducted electrical and optical
three-signal tests so far are at least an order of magnitude above the
required precision, mainly limited by
testbed noise \cite{Mitryk2010,Devine2010}. Digital
three-signal tests \cite{Shaddock2006,Gerberding2013} verified
performance of the digital phasemeter core (presented later), while by
definition they cannot test the typically limiting analog front-end.

This Letter reports on a hexagonal quasimonolithic optical bench
implementing an optical three-signal test. It shows sufficient stability for LISA phasemeter
linearity and precision verifications. After the experimental setup, measurements
proving the capabilities of the testbed together with a benchmark
performance test of a LISA phasemeter will be shown. 

\textit{Experimental setup}---Firstly, the optical three-signal scheme will
be described. Here, phase is defined as instantaneous phase $\varphi(t)$ of
a harmonic signal $x(t)$ with amplitude $a$. It is closely related to the 
instantaneous frequency $f(t)$ and the carrier frequency offset $\overline{f}$: 
$$x(t)=a\cdot\sin(\varphi(t)),  \qquad d \varphi (t) / dt
=2\pi (f(t)+\overline{f}).   $$  
For testing, three initial signals (electric fields of laser beams
exhibiting unequal optical frequency offsets) 
with phases $\varphi_1$, $\varphi_2$, $\varphi_3$ are interfered
pairwise. Each resulting laser beam intensity contains two mixed
initial signals which generate a beat note. The beat note signals can exhibit
unequal MHz frequency offsets, called heterodyne frequencies $\overline{f}_a\neq
\overline{f}_b \neq \overline{f}_c$, and phases
$\varphi_a$, $\varphi_b$, $\varphi_c$:
$$\varphi_a=\varphi_1-\varphi_2, \quad \varphi_b=\varphi_2-\varphi_3, \quad 
\varphi_c=\varphi_3-\varphi_1. $$ 
After conversion to voltages by photoreceivers, three phasemeter channels extract the measured phases $\varphi_a'$,
$\varphi_b'$ and $\varphi_c'$ from the beat notes. With the operator
$\mathcal{E}$ denoting the phase fidelity of the extraction, it can be
written:
$$ \varphi_a' = \mathcal{E}(\varphi_1 - \varphi_2), \quad \varphi_b' =
 \mathcal{E}(\varphi_2 - \varphi_3), \quad \varphi_c' = \mathcal{E}(\varphi_3
 - \varphi_1). $$ 
Finally, the three measured phases are combined in postprocessing to form the
three-signal measurement
$$\varphi_0=\varphi_a'+\varphi_b'+\varphi_c' \stackrel{?}{\sim}0,$$
which is the main measurand and in which the initial phases ideally
cancel. It includes the phase noise contribution of the phasemeter
 while being sensitive to nonlinearities: if $\mathcal{E}$ is nonlinear, which means the condition
$\mathcal{E}(\varphi_1-\varphi_2)=\mathcal{E}(\varphi_1)-\mathcal{E}(\varphi_2)$
does not hold, the generic initial phases will not cancel pairwise in
$\varphi_0$. The same is true for nonlinear effects due to the
unequal heterodyne frequencies. Additionally, the ratio between $\varphi_0$
and the single channel inputs $\varphi_{a-c}'$ gives a direct estimate of
the phasemeter dynamic range. 

Here, the scheme is implemented using
a hexagonal quasimonolithic bench as testbed core. The complete testbed is divided into three conceptual parts.

The first part is a laser preparation bench. It provides the initial signals with phases
$\varphi_{1-3}$ using three 1064 nm lasers (Mephisto by Coherent, 500 mW).  As their pairwise combinations
are supposed to generate beat notes with heterodyne frequencies of 5-25 MHz,
their frequency relation must be well-defined. This is achieved by two
digital control loops locking the frequencies of two slave lasers to one
master laser. The loop reference signal then sets the desired heterodyne
frequencies and can be utilized to add artificial LISA-like
frequency noise to $\varphi_{a-c}$ to test the required dynamic range.

The second part is the LISA phasemeter under test here. It was developed within an
ESA contract \cite{Barke2014} by a European consortium. It is based on
parallel ADC channels (80 MHz) connected to field-programmable
gate arrays (FPGAs) implementing digital phase-locked loops (DPLLs). 
The frequencies tracked by the DPLLs are downsampled and converted to
phase in postprocessing.  While the performance of the DPLL serving as phasemeter core was
verified in the aforementioned digital three-signal test \cite{Gerberding2013}, it was also shown that the
phasemeter utilized here fulfills the LISA requirements in electrical split tests \cite{Gerberding2015}.

\begin{figure}[bp]
 \includegraphics{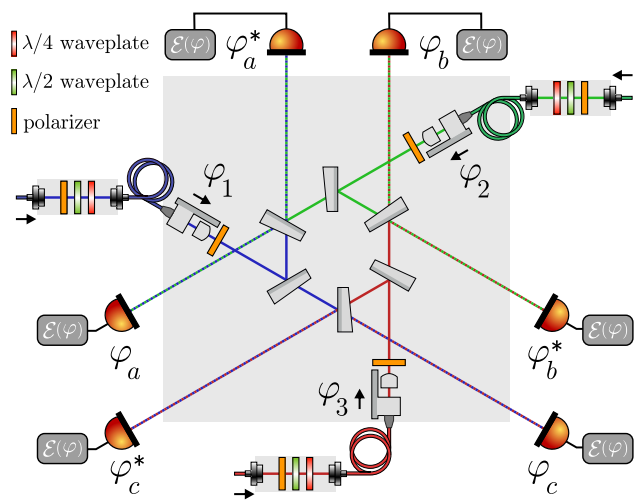}
 \caption{\label{FigHex}Schematic of the hexagonal quasimonolithic optical
   bench implementing a three-signal linearity test. Complementary
   interferometer outputs are utilized for diagnostic $\pi$ pretests.}
\end{figure}

Finally, the core of the experiment is the aforementioned quasimonolithic optical bench. It was
designed using the software tool IfoCAD \cite{Wanner2012} and enables the
stable splitting and interference of the initial signals according to the
three-signal scheme. Fig. \ref{FigHex} shows a schematic of the bench. It consists of a
thick Zerodur baseplate ($200 \times 200 \times 25$ mm) carrying three
fiber injector optical subassemblies (FIOSs) for the
injection of the prepared laser light and six wedged beam splitters
(roughly 50:50 split ratio) placed in a
hexagonal layout.  Three of the
beam splitters act as a first stage where each of the
injected beams (carrying the phases $\varphi_{1-3}$) is divided into
two.  In a second stage, each divided beam is interfered
pairwise with the output of another dividing beam splitter, forming 
three interferometers in total.  Finally, the
interferometer outputs at the
second stage beam splitters yield the three beat notes with
heterodyne frequencies $\overline{f}_{a-c}$ and phases $\varphi_{a-c}$ as well as three
complementary versions with phases $\varphi_{a-c}^*$ (shifted by
$\pi$). The six output beat notes are captured by
 photoreceivers and are subsequently sampled by the
phasemeter. The photoreceivers, designed for the testbed using
off-the-shelf components, comprise InGaAs photodiodes (0.5 mm in diameter,
$\sim$1 mW incident power) and transimpedance amplifiers with a single
operational amplifier. At each output beam splitter, one of the
complementary signals can be picked for the three-signal test. Also, both
signals' phases can be averaged for balanced detection or be subtracted for
an optical split test with a $\pi$ phase shift, named $\pi$
measurement. The latter exhibits shortcomings similar to the split tests
described above, and is considered a diagnostic pretest. It can reveal noise
sources it has in common with the three-signal test, e.g. some types of
stray light. A subset of these can be canceled with balanced detection.

While not being the only topology possible for a three-signal scheme, the
presented approach is well-suited concerning two
critical aspects. One is the phase noise coupling after the split of the
initial phases. 
In general, this kind of noise $\varphi_{\mathrm{N}}$ is not common and acts in one
phasemeter channel only, as e.g. $\mathcal{E}(\varphi_1-\varphi_2+\varphi_{\mathrm{N}})$
shows. Hence, it cannot be distinguished from phasemeter noise and limits the testbed.
In the hexagonal bench, this noise is primarily determined by the displacement stability between the first and second stage
beam splitters.  The hexagonal configuration allows a compact and
symmetric implementation of the interferometers,
hence lowering the noncommon-mode displacement fluctuations caused by thermal expansion and mechanical
distortions. For further mitigation of those effects, the thick Zerodur
baseplate serves as thermal
bulk with low thermal expansion coefficient while the fused silica beam splitters are
attached via hydroxide-catalysis bonding \cite{Elliffe2015} and optical
contacting. Due to its repeatability, the latter method was chosen for the
placement of the second stage beam splitters which are decisive for proper contrast.

A second critical aspect of the bench is the static mismatch of the
two displacements between any first stage beam splitter and its successive
second stage beam splitters. In general, such a mismatch acts
as unequal interferometer arms and hence leads to the coupling of single laser
frequency noise limiting the testbed. This unwanted noise should not be
confused with the controlled differential laser frequency noise in $\varphi_{a-c}$ used to mimic the
master laser frequency noise in LISA and which is meant to cancel out. Nevertheless, the symmetric layout
of the hexagonal interferometers allows the reduction of the coupling
by matching the static displacements within assembly tolerances. A maximum
displacement mismatch of $\sim 200 \ \mu  \mathrm{m}$ is assumed.

As an amendment to the stable hexagonal interferometer design, the
aforementioned FIOSs were added to minimize thermally induced angular jitter of the input
beams.  The FIOSs consist of glued fused silica
components and are based on adaptations of earlier designs \cite{Killow2016}.  The
minimization of the angular jitter is desirable as it couples into
displacement noise via the wedged beam splitters. The latter in turn were
chosen to achieve the angular separation of desired beams and ghost beams reflected from secondary surfaces.

The optical bench together with auxiliary optics and photoreceivers is placed in a vacuum chamber. For proper operation,
a moderate vacuum (roughly $<10\ \mathrm{mbar}$) is required, primarily to avoid optical path length
fluctuations caused by residual air. A fiber interface connects the
external laser preparation to the optical bench.  

Another essential aspect for the operation of the testbed is a proper
polarization control. Ideally, all beams in the interferometer should
exhibit the same polarization axis. Mismatches lead to secondary parasitic
interferometers in the orthogonal axis. The closest approach to
the ideal case was achieved with laminated thin-film polarizers (extinction ratio  $1{:}10^7$) placed right after
the FIOSs and with their transmissive axis set to the bench surface normal as a
common reference. The polarization cleanliness was improved 
further by controlling the input of the polarization-maintaining FIOS fibers. For that purpose,
pairs of $\lambda/2$- and $\lambda/4$-waveplates optimally match the fiber
input beam polarization to the fiber slow axis. This also attenuates indirect
coupling chains, e.g. fiber polarization fluctuations to pointing
jitter to phasefront jitter to phase noise.

To summarize, the presented setup aims to minimize all testbed
noise indistinguishable from the noise contribution of the
phasemeter. The residual noise floor gives an upper bound for the
phasemeter performance in a three-signal test, while the digital
laser control can create LISA-like phase input
conditions.

An important extension will be the utilization of three separate phasemeters with
independent clocks \cite{Schwarze2016}, which would allow testing of
LISA SC-SC features like clock tone transfer, ranging and data transfer as well as postprocessing techniques
like interpolation, clock synchronization and clock noise
removal for TDI. This way, the setup can provide realistic data to support scientists developing the LISA data
processing and analysis. Also, the attenuation of a single beam to 100 pW
will allow proper testing of dedicated LISA photoreceivers.

\begin{figure}[tbp]
 \includegraphics{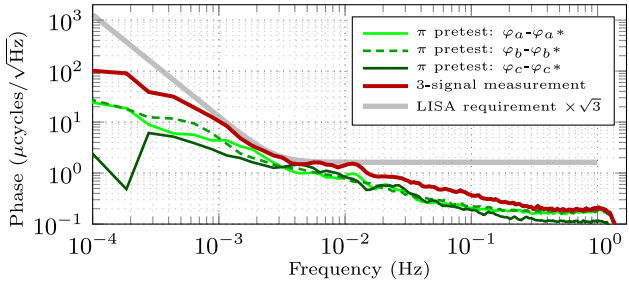}%
 \caption{\label{Meas1}Measurement (3 h) with moderate phase noise input 
   conditions. The three-signal combination (red) fulfills the LISA requirement
   scaled for three signals. Its fundamental noise can be found in the
  $\pi$ pretests (green), suggesting sufficient stability of the
  hexagonal optical bench itself.}
\end{figure}

\textit{Results}--- In the following, three measurements carried out with the described setup are
presented. One was conducted with low heterodyne frequencies (5.8 MHz, 3.01
MHz, 2.79 MHz) and input phase noise of $0.04 \
\mathrm{cycles}/\sqrt{\mathrm{Hz}}$ at 1 Hz. These values allowed
excluding from the performance assessment effects such as dynamic range
limitations or noise caused by high carrier frequencies. The resulting
spectral densities \cite{Troebs2006,Troebs2009} are shown
in Fig. \ref{Meas1}.  The three-signal performance (red line, best pick from
each of the complementary output pairs) satisfies the
LISA phase extraction requirement scaled for three uncorrelated
signals. Above 1 mHz, the $\pi$ pretests (green lines) at the three interferometer
outputs show a similar noise shape and magnitude, while balanced
detections (not shown) did not yield a significant improvement. Hence it is assumed
that, in this frequency band, the dominant noise source in the three-signal test is
the same as in the $\pi$ pretests. Vanishing as common-mode in the latter, instability of the
optical bench can be excluded. Instead, residual polarization mismatches
are considered as the main limitation candidate, as small changes in the polarizer alignments easily
spoil both $\pi$ and three-signal tests.  This suggests that the testbed performance, not limited
by the hexagonal optical bench itself, could be improved further.

\begin{figure}[tbp]
 \includegraphics{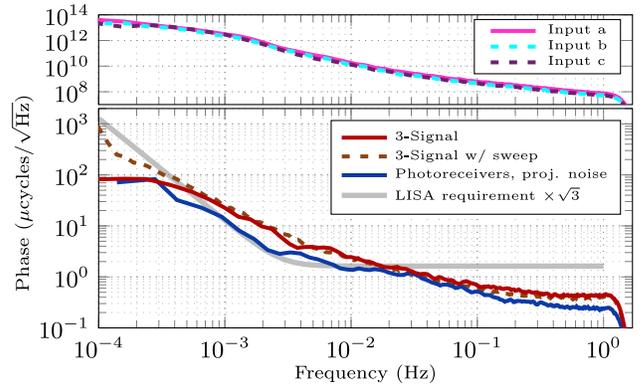}%
 \caption{\label{Meas2}Measurements with LISA-like input phase noise and
   heterodyne frequencies. A dynamic range of 8-11 orders of
   magnitude can be computed from the three input signals (top) and
   the three-signal combinations (red: fixed heterodyne frequencies, 3 h; brown
   dashed: heterodyne frequency sweep over 90 h). The LISA
   three-signal requirement was fulfilled except between 0.4-20 mHz, with
   the photoreceivers being a major noise contributor (noise projection in blue).}
\end{figure}

\begin{figure}[tbp]
 \includegraphics{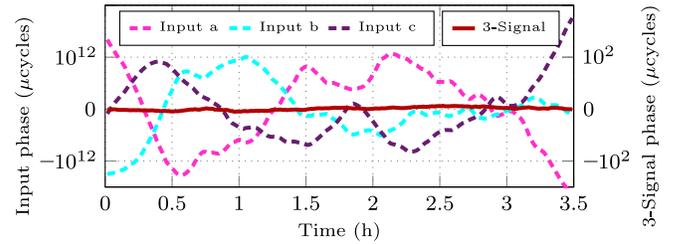}%
 \caption{\label{Meas2_TS}Time series of input phase fluctuations and resulting
   three-signal combination, illustrating the high dynamic range essential
   for TDI.}
\end{figure}

More measurements were conducted with LISA-like input conditions as specified in the following. The results are shown in
Fig. \ref{Meas2}. For the first three-signal measurement (red solid line), the heterodyne frequencies
were set to 24.9 MHz, 18.1 MHz and 6.8 MHz, while the single input phases (top lines) were generated to resemble
a LISA-like signal shape with instantaneous frequency noise
of $450 \ \mathrm{Hz} /\sqrt{\mathrm{Hz}}$ at 1 Hz ($\sim 70 \ \mathrm{cycles}/\sqrt{\mathrm{Hz}}$)
and a $1/f$ behavior dominating below 3 mHz.
With the shown three-signal performance, this corresponds to a dynamic
range of 8, 10 and 11 orders of magnitude at 1 Hz, 3 mHz and below 1 mHz,
respectively. This is illustrated in Fig. \ref{Meas2_TS} which shows time series of the single input 
phase fluctuations and their combination in a drastically reduced scale (right side). The
performance satisfied the LISA phase extraction requirement except in
the range of 0.4-20 mHz where a violation by a factor 3 occurs. Nevertheless, when summed
quadratically with other noise sources for the overall $10 \
\mathrm{pm} /\sqrt{\mathrm{Hz}}$ budget, it is still not a significantly limiting contribution. 
A major contribution to the higher noise level compared to the measurement with moderate input
conditions was traced back to the utilized photoreceivers. At the required
precision levels, they show a heterodyne-frequency-dependent noise behavior and thus lead to excess
noise when operated at the upper end of the LISA heterodyne frequency band.
Isolated differential measurements between pairs of photoreceivers were
conducted for a noise projection (Fig. \ref{Meas2}, blue line). This
includes a higher noise contribution by the phasemeter itself, which shows
a similar, but weaker heterodyne-frequency dependence.


A second long-term three-signal measurement (Fig. \ref{Meas2}, brown dashed line) was conducted with similar input
phase noise as in the prior measurement but with dynamic heterodyne
frequencies.  More specifically, sweeps from 24.9-5.0 MHz, 18.1-3.0 MHz, and
6.8-2.0 MHz were applied within a time frame of 90 h. This corresponds to a maximum
sweep rate of $\sim$ 61 Hz/s. Compared to the prior measurement, the
performance shows no significant deterioration except a stronger low-frequency
drift. 

\textit{Discussion and conclusions}--- The measurement presented in
Fig. {\ref{Meas1}} shows that the hexagonal
optical bench provides sufficient stability down to LISA-like picometer
levels and a static displacement matching that allowed the use of a free-running master laser. These properties
in turn enabled the measurement to be the first optical three-signal linearity test with MHz
signals and $\mu \mathrm{cycle}/\sqrt{\mathrm{Hz}}$ precision. Additionally, the $\pi$ pretests
suggest that the hexagonal bench stability is currently not limiting above
1 mHz and that the performance could be improved further. 

Nevertheless, the testbed already enabled a benchmark linearity
test of a phasemeter with LISA-like dynamic range and heterodyne
frequencies (Fig. \ref{Meas2}). The required phase extraction performance could be verified in most of the
frequency band, with the phasemeter likely not being limiting in the
current state. Instead, a major noise contributor are the photoreceivers.
Yet, they are not considered a show stopper and will be investigated further.

Comparable state-of-the-art testbeds \cite{Devine2010,Mitryk2012} were able to show
similar dynamic ranges, however with single-digit MHz frequencies and most
importantly with precision levels more than an order of magnitude above the
ones demonstrated in this Letter.

The shown results suggest the utilization of the hexagonal optical bench testbed as a
facility for the verification of future iterations of the LISA
phasemeter, including engineering or flight models. As mentioned, the testbed can easily
be extended to probe other important features of the LISA metrology chain
and to support LISA data processing and analysis. 

Besides this extension, future work will include efforts to tackle the noise
sources assumed to be limiting, like polarization and photoreceivers, to reduce the
testbed noise floor further.

To conclude, the LISA phasemeter in particular and the LISA metrology chain
in general are crucial for the successful detection of gravitational
waves in space. Stringent requirements are imposed on these components,
making verification a challenge. The measurements presented here show
that the hexagonal optical bench provides the capability to face this challenge successfully.

\begin{acknowledgments}
The authors acknowledge financial support by the European Space Agency
(ESA) (22331/09/NL/HB, 16238/10/NL/HB), the German Aerospace Center (DLR)
(50OQ0601, 50OQ1301) and the Sonderforschungsbereich (SFB) 1128
Relativistic Geodesy and Gravimetry with Quantum Sensors (geo-Q).
\end{acknowledgments}

\end{document}